\documentclass[conference]{IEEEtran}
\IEEEoverridecommandlockouts
\usepackage{cite}
\usepackage{amsmath,amssymb,amsfonts}
\usepackage{algorithmic}
\usepackage{graphicx}
\usepackage{textcomp}
\usepackage{xcolor}
\usepackage{balance}
\def\BibTeX{{\rm B\kern-.05em{\sc i\kern-.025em b}\kern-.08em
    T\kern-.1667em\lower.7ex\hbox{E}\kern-.125emX}}

\newcommand{\telechat}{T\'el\'echat}

\usepackage{float}
\usepackage{listings}
\lstdefinestyle{mystyle}{
    basicstyle=\ttfamily\footnotesize,
    breakatwhitespace=false,         
    breaklines=false,                 
    captionpos=b,                    
    keepspaces=true,                 
    numbers=left,                    
    numbersep=5pt,                  
    showspaces=false,                
    showstringspaces=false,
    showtabs=false,                  
    tabsize=2,
    numbers=none
}

\usepackage{pifont}
\newcommand{\cmark}{\ding{51}}%
\newcommand{\xmark}{\ding{55}}%

\usepackage{tikz}
\usepackage{tikz-cd}
\usepackage{mathtools}
\usepackage{verbatim}

\usepackage{amsfonts}
\usepackage{amssymb}

\RequirePackage[hyphens]{url}
\usepackage[hidelinks]{hyperref}

\usepackage{amsthm}

\theoremstyle{definition}
\newtheorem{definition}{Definition}[section]

\usepackage{microtype}

\begin{document}

\title{Compiler Testing with Relaxed Memory Models}

\author{\IEEEauthorblockN{Luke Geeson}
\textit{University College London, UK}\\
\url{luke.geeson@cs.ucl.ac.uk}
\and
\IEEEauthorblockN{Lee Smith}
\textit{Arm Ltd*\thanks{*Smith retired from Arm at the end of 2022.}, UK}\\
\url{lee.d.smith@acm.org}
}

\maketitle
\thispagestyle{empty}
\pagestyle{empty}

\begin{abstract}
Finding bugs is key to the correctness of compilers in wide use today. If the behaviour of a compiled program, as allowed by its architecture memory model, is not a behaviour of the source program under its source model, then there is a bug. This holds for all programs, but we focus on concurrency bugs that occur only with two or more threads of execution. We focus on testing techniques that detect such bugs in C/C++ compilers.

We seek a testing technique that automatically covers concurrency bugs up to fixed bounds on program sizes and that scales to find bugs in compiled programs with many lines of code. Otherwise, a testing technique can miss bugs. Unfortunately, the state-of-the-art techniques are yet to satisfy all of these properties.

We present the \telechat{} compiler testing tool for concurrent programs. \telechat{} compiles a concurrent C/C++ program and compares source and compiled program behaviours using source and architecture memory models. We make three claims: \telechat{} improves the state-of-the-art at finding bugs in code generation for multi-threaded execution, it is the first public description of a compiler testing tool for concurrency that is deployed in industry, and it is the first tool that takes a significant step towards the desired properties. We provide experimental evidence suggesting \telechat{} finds bugs missed by other state-of-the-art techniques, case studies indicating that \telechat{} satisfies the properties, and reports of our experience deploying \telechat{} in industry regression testing.
\end{abstract}


\section{Introduction}\label{sec:intro}

Finding compiled program behaviours, or \textit{bugs}, that are forbidden by the source program's language semantics, is key to ensuring compiler correctness. Finding concurrency bugs in compilers is especially important, as more programs are compiled for multicore processors each year. Unfortunately, finding such bugs can be tricky, as concurrent programs exhibit behaviours that can be unintuitive. To complicate matters, multi-core processors may execute each thread of a concurrent program \textit{out-of-order}, influencing the execution of other threads through shared memory. This is \textit{relaxed memory concurrency}, and is exhibited by processors based on the Arm architecture, Intel x86-64, IBM PowerPC, RISC-V, MIPS, and more. As such concurrency bugs can require conditions that rarely occur in practice. We address the problem of \textit{how} to find concurrency bugs introduced by compilers when preparing programs for these architectures.
\begin{figure}[t]
\centering
  \begin{tabular}{c}
\begin{lstlisting}[language=C, style=mystyle]
{ *x = 0; *y = 0; } // fixed initial state

#define relaxed memory_order_relaxed
#define release memory_order_release
#define acquire memory_order_acquire

// Concurrent Program with threads
P0 (atomic_int* y,atomic_int* x) {
  atomic_store_explicit(x,1,relaxed);
  atomic_thread_fence(release);
  atomic_store_explicit(y,1,relaxed);
}
P1 (atomic_int* y,atomic_int* x) {
  atomic_exchange_explicit(y,2,release);
  atomic_thread_fence(acquire);
  int r0 = atomic_load_explicit (x,relaxed);
}

// Predicate over the final state
exists (P1:r0=0 /\ y=2)
\end{lstlisting}
  \end{tabular}
  \caption{Litmus tests have a fixed initial state, a concurrent program and a predicate over the final state. Outcomes that satisfy the \texttt{exists} clause are forbidden by the C/C++ model~\cite{C11}. When compiled the outcome is allowed by the Armv8 AArch64~\cite{aarch64} model. We found this bug~\cite{swp} using \telechat{}.}
  \label{lb_rc11}
\end{figure}

For a compiler to be deemed correct, the compiled program must behave as allowed by the semantics of its source~\cite{Leroy:2009:FVR:1538788.1538814}. The behaviour of a concurrent program can be defined by its set of \textit{executions} - characterised by the communications between threads of execution through shared memory~\cite{10.1145/2487241.2487248}. A \textit{memory consistency model}~\cite{Alglave:2014:HCM:2633904.2627752}, such as the ISO C/C++~\cite{C11} model $M_{S}$, describes the set of allowed executions of a C/C++ source program $S$. Likewise, the Armv8 model $M_C$ in \textsection B2.3.1 in the Arm Architecture Reference Manual~\cite{Seal:2000:AAR:517257} describes the allowed executions of a compiled program $comp(S)$. All widely used processor architectures have published memory models. A correct compiler must ensure that the allowed source program executions include the allowed compiled program executions for each well-defined concurrent program $S$:
\begin{multline*}\label{eq:correct}\tag{eq.1}
\forall S. outcomes(exec(comp(S),M_{C})) \subseteq \\ outcomes(exec(S,M_{S}))
\end{multline*}
If the outcomes of source program executions do not include the outcomes of compiled executions under each respective model ($exec(P,M)$ runs $P$ under a model $M$), then there is a bug. Of course, this holds for all programs and so we focus on bugs that are observable only with two or more threads.

We are motivated to test production compilers developed by Arm's engineers. Such compilers can undergo many revisions each day for which the repeated formal proof of \ref{eq:correct} is infeasible~\cite{10.1145/3460319.3469079}. Comparing unbounded executions under relaxed memory is also undecidable~\cite{10.1145/1706299.1706303}. Instead, we conduct bounded \textit{testing}. We assist Arm's compiler teams who wish to deploy automated compiler testing for concurrent C/C++.

We seek a technique with four properties. Firstly, a technique needs \textit{coverage} of bugs up to fixed bounds on programs with a fixed initial state, loop unroll factor, and no recursion. A technique without coverage may miss bugs and cannot be reliably deployed in automated testing. Second, a \textit{general} technique should support current and future models of each source language and assembly language supported by the compiler under test, else it can miss bugs as architecture specifications evolve. Thirdly, a technique should find bugs in given tests \textit{automatically} - without further input. Of course, finding concurrency bugs can take days, which makes testing daily compiler revisions impractical. Testing should therefore \textit{scale} to find bugs in programs with many threads and lines of code (LoC) per thread \textit{quickly} (two minutes). Without these properties a technique can miss bugs, as we require a reliable and repeatable means of testing each compiler revision.

Unfortunately, the state-of-the-art tools~\cite{cmmtest,Chakraborty,10.1145/3460319.3469079} are yet to satisfy the four properties. The \texttt{C4} tool~\cite{10.1145/3460319.3469079,9477685,windsor2022high} exploits the scalability of hardware, but hardware may miss bugs~\cite{windsor2022high}, as bugs can occur perhaps once in thousands of runs of a compiled program, if hardware implements the required behaviour at all. \texttt{validc}~\cite{Chakraborty} and \texttt{cmmtest}~\cite{cmmtest} compare all bounded executions, but require experts to find the bugs. As far as we know, these works are not deployed in industry.

We present the \telechat{} compiler testing tool. Given a C/C++ program, \telechat{} prepares source and compiled programs for testing, using the \texttt{herd}~\cite{Alglave:2014:HCM:2633904.2627752} simulator. \telechat{} finds bugs when there is an outcome of executing the compiled program under the architecture model that is not an outcome of executing the source program under the source model.

We claim that \telechat{} is the first tool to satisfy the four properties. By using the \texttt{herd} simulator, \telechat{} finds bugs \textit{automatically}. By relying on official models, \telechat{} \textit{covers} the behaviour allowed by authoritative C/C++ and architecture standards. Coverage is \textit{general}, since we parameterise over both source and target models. Our technique \textit{scales}, since \telechat{} optimises compiled programs. Significant work was required to make testing scale, as \texttt{herd} is designed to test small programs, and execution time expands factorially as the test size increases, practically limiting its ability to scale much above programs of the order of 40-50 LoC. \telechat{} makes significant steps towards scalable compiler testing in practice as checking compiled programs terminates in milliseconds. 

\telechat{} improves on the state-of-the-art for the task of finding C/C++ concurrency bugs. In other words, the set of bugs found by the state-of-the-art are a subset of bugs found by \telechat{}. We contribute experimental evidence that suggests \telechat{} finds behaviours missed by the state-of-the-art on the same inputs and models. As far as we know, \telechat{} is the first publicly available compiler testing tool (for concurrency) to be deployed in industry.

The rest of this work is structured as follows. \textsection\ref{sec:bg} covers the background and literature review. \textsection\ref{sec:wip} covers the design, implementation, reproducible artefact, and documentation for \telechat{}. We evaluate the efficacy of \telechat{} in \textsection\ref{sec:eval} and conclude in \textsection\ref{conc} with lines of inquiry exposed by our work.

\subsection{Our Contributions}

\textbf{Technique, Tool, and Artefact}
\begin{itemize}
\item Novel compiler testing technique parameterised over source and architecture memory models.
\item The \telechat{} tool that implements our technique.
\item An artefact is available to reproduce experiments using benchmark tests and documentation. 
\end{itemize}

\textbf{Bug-Finding Campaign}
\begin{itemize}
\item Three new compiler bugs: Reported a run-time crash~\cite{constbug} induced by \texttt{const}-qualified atomic loads in LLVM for the Armv8 architecture, a wrong-endian bug~\cite{stp} in the compilation of 128-bit atomic store instructions, (Armv8) a concurrency bug~\cite{ldpbug} in the compilation of 128-bit sequentially consistent~\cite{Lamport:1979:MMC:1311099.1311750} loads (Armv8), and an optimisation opportunity~\cite{mipbug} in the GCC MIPS backend.
\item A new model bug: Fixed a bug~\cite{dmbish} in the unofficial Armv7 model that allowed behaviour forbidden on Arm hardware.
\item One new bug type: Refute a claim made by Morisset et. al~\cite{cmmtest} and identify a new kind of bug~\cite{swp} (Fig.~\ref{lb_rc11}).
\end{itemize}

\textbf{Controlled Experiments}
\begin{itemize}
\item Found a concurrency behaviour (Fig.~\ref{lb_c4}) known to experts but missed by the state-of-the-art \texttt{C4}~\cite{10.1145/3460319.3469079,windsor2022high}.
\item Found two concurrency behaviours: Conducted large-scale differential testing of LLVM and GCC for Armv8, Armv7, Intel, RISC-V, PowerPC, and MIPS architectures. With 9 million tests, it is the most extensive concurrency test campaign as far as we know.
\end{itemize}

\textbf{Industry Experience}
\begin{itemize}
\item Addressed a practical limitation of simulation. \texttt{herd} was designed to simulate \textit{small} tests and many authors~\cite{cmmtest,10.1145/3563292,inbook} claim it is unlikely to scale to finding bugs using large tests. We optimise compiled programs and \texttt{herd} runs much faster, often terminating in milliseconds.
\item Answered queries from Arm's partners~\cite{LDAPR} concerning \texttt{LDAPR} and \texttt{LDAR} instructions.
\item Deployed \telechat{} in automated testing for Arm Compiler. As far as we know, \telechat{} is the industry's first publicly available technique that is deployed in automated testing, and has tested Arm Compiler since June 2022.
\end{itemize}

\subsection{\telechat{} Benefits}
\begin{itemize}
\item \textbf{Futureproof}, \telechat{} tests compilers against architectural memory models, and those memory models are typically designed to describe the limits of permissible orderings of the architecture, including permissible ordering behaviours that will only be seen on future hardware or on hardware that is not readily accessible, for example Morello~\cite{morello}.
\item \textbf{Familiar} to engineers who are not necessarily concurrency experts. Arm's engineers are using litmus tests to discuss concurrency queries as they arise.
\item \textbf{Authoritative} oracle. By using official architectural models, \telechat{} approaches a ground truth for compiler testing - reducing the bug-finding problem to test generation. 
\end{itemize}

\section{Preliminaries}\label{sec:bg}
We illustrate the concepts involved using the example bug report~\cite{swp} in Fig~\ref{lb_rc11}, known as \textit{message passing}.

\subsection{Litmus Tests and Memory Models} \label{tools}
\begin{figure}[h]
\centering
  \begin{tabular}{l l}
    \includegraphics[scale=0.6, trim={1cm 0 1cm 0}]{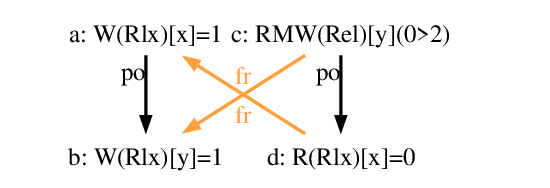} &
    \includegraphics[scale=0.6, trim={1cm 0 1cm 0}]{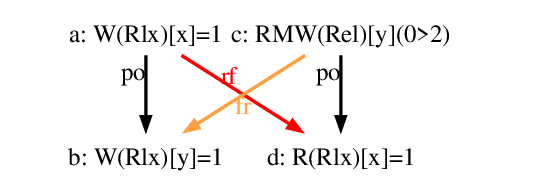} \\
    \includegraphics[scale=0.6, trim={1cm 0 1cm 0}]{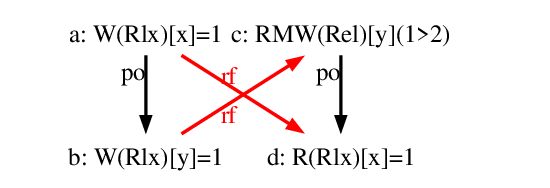} &
    \includegraphics[scale=0.6, trim={1cm 0 1cm 0}]{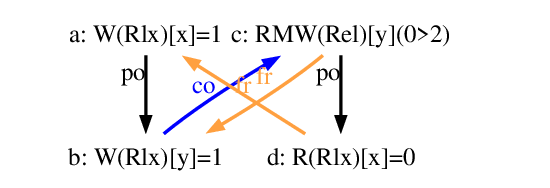} \\  
  \end{tabular}
  \caption{Executions of Fig.~\ref{lb_rc11}, \texttt{a:R(Rlx)[x]=0} means event $a$ reads the value 0 from location $x$ with relaxed memory ordering. Top left is \textbf{acbd} or \textbf{cabd}, right \textbf{abcd}, bottom left \textbf{cdab}, right \textbf{dabc} (\textbf{dabc} is forbidden by RC11~\cite{simonrc11}).}
  \label{lb_exe}
\end{figure}

\textit{Litmus tests} - like in Fig.~\ref{lb_rc11} - are used to explore executions allowed by hardware or a model. Litmus tests define a fixed initial state, a concurrent program, and a predicate over the final state. A concurrent program defines multiple threads (\texttt{thd}=\texttt{P0},\texttt{P1},\ldots) that read from or write (\textit{events} \texttt{E}, in the terminology of \textsection B.2 of~\cite{Seal:2000:AAR:517257}) to shared memory locations (\texttt{loc=x,y,z,}\ldots). When threads communicate, they produce one or more \textit{executions} as shown in Fig.~\ref{lb_exe}. The \texttt{diy} tool~\cite{Alglave:2012:FWM:2205506.2205523} generates litmus tests from executions, such as the execution \textbf{dabc} in Fig.~\ref{lb_exe}. Isla~\cite{isla-cav}, Dartagnan~\cite{10.1145/3563292}, and Memalloy~\cite{Wickerson:2017:ACM:3009837.3009838} use SMT solvers to explore executions in a similar fashion.

\begin{definition}{\textit{Execution}}:\label{exe} A graph where nodes are events and edges are partial order relations over events~\cite{Alglave2021ArmedCF}. An execution is \textit{allowed} if it is exhibited by hardware or a model, else it is \textit{forbidden}. The base relations are:
\begin{itemize}
\item \textit{program-order (po)}$\triangleq$\{ (\texttt{E}$_{1}$,\texttt{E}$_{2}$)\texttt{|thd(E$_{1}$)=thd(E$_{2}$)} $\land$\texttt{E$_{1}$};\texttt{E}$_{2}$\} \textit{ie} the order instructions are written on the page
\item \textit{reads-from (rf)} $\triangleq$ \{ \texttt{(W,R)| loc(W)=loc(R) $\land$ val(W) = val(R)} \}
\item \textit{coherence (co)}$\triangleq$\{ (\texttt{W$_{1}$,W$_{2}$)|loc(W$_{1}$)=loc(W$_{2}$)} \}
\item \textit{from-read (fr)} $\triangleq$ \{(\texttt{R},\texttt{W}$_{1}$) \texttt{| $\exists\,$W$_{2}$}. \texttt{W}$_{2} \xrightarrow{rf}$ \texttt{R} $\land$ \texttt{W}$_{2} \xrightarrow{co}$ \texttt{W}$_{1}$\}
\end{itemize}
\end{definition}

Executions abstract machine operations as mathematical objects. Executions model architecture features, such as pipelines or caches, as their effects on the order that reads and writes reach shared memory. For a given thread, the order that its accesses reach memory influences the executions of other threads, which influence memory in turn. When a litmus test is run from some fixed initial state, the interaction between all threads of execution produces a set of \textit{candidate} executions. 

A \textit{memory consistency model} (or a model) filters out forbidden executions of a litmus test. Models define predicates on relations over events, forbidding - for instance - cyclic executions.  The Cat~\cite{DBLP:journals/corr/AlglaveCM16} language specifies models of RC11~\cite{simonrc11}, Armv8 AArch64~\cite{aarch64} (official), Armv7~\cite{aarch32} (unofficial), RISC-V~\cite{lucrisc} (official), Linux~\cite{Alglave:2018:FSC:3173162.3177156}, Intel x86-64~\cite{x86tso}, IBM PowerPC~\cite{ppc}, MIPS~\cite{mips}, and more. The \texttt{herd} \textit{simulator}~\cite{Alglave:2014:HCM:2633904.2627752} enumerates the executions of small litmus tests (from fixed initial states up to fixed loop unroll factor with no recursion) allowed by a Cat model. Fig.~\ref{lb_log} shows the \textit{outcomes} of the executions of Fig.~\ref{lb_rc11} allowed by the RC11 C/C++ model~\cite{simonrc11}.

\begin{definition}{\textit{Outcome}}: \label{out} An outcome is the result of an execution (def.~\ref{exe}) expressed as a set of assignments to shared memory (e.g. \textnormal{\texttt{y=2}}) and thread-local data (e.g. \textnormal{\texttt{P1:r0=1}}). The set of outcomes of executing a litmus test $P$ is denoted $outcomes_P$. We defer other effects (like IO) to future work.
\end{definition}
\begin{figure}[H]
\centering
  \begin{tabular}{c}
\begin{lstlisting}[language=C, style=mystyle]
{ P1:r0=0; [y]=1; } // outcome acdb of Fig.1
{ P1:r0=1; [y]=1; } // abcd
{ P1:r0=1; [y]=2; } // cdab
\end{lstlisting}
  \end{tabular}
  \caption{The outcomes of executions in Fig.~\ref{lb_exe}. The \texttt{acyclic} constraint of the RC11 model~\cite{simonrc11} forbids \textbf{dabc} and its outcome \texttt{\{P1:r0=0; y=2\}}.}
  \label{lb_log}
\end{figure}
The Arm AArch64 and RISC-V models are maintained by their respective architecture specification teams. Other models used are from peer-reviewed publications. We build on these models and rely on their correctness.

The \texttt{litmus} tool~\cite{Alglave:2011:LRT:1987389.1987395} runs litmus tests on hardware to check if hardware correctly implements models. If hardware exhibits forbidden executions then either the model is wrong, or the hardware is incorrectly implemented. As silicon manufacturers may implement restricted variants of an architecture model, hardware executions may omit behaviours allowed by the model. The \texttt{litmus} tool is therefore of limited use to compiler testing. To be clear, it \textit{is} necessary to validate hardware against models but that is a \textit{separate} problem from validating \textit{compilers} against models.

Tools that use hardware as an oracle for correct behaviour are unlikely to be reliable for compiler testing. Observing behaviours on hardware may require circumstances that rarely occur in practice, if at all. The chances of observing a behaviour depend on whether a given implementation supports it and whether the hardware is in a sufficiently stressed state. Observations may require sampling a vast array of hardware many (thousands of) times to reliably test a compiler revision. Even then, testing on hardware may miss bugs.

\subsection{Compiler Testing}\label{testing}
We use the testing terminology from Barr et al.~\cite{10.1109/TSE.2014.2372785}. To test is to stimulate a system under test and observe its response~\cite{10.1109/TSE.2014.2372785}. We stimulate the system under test $comp$, with a source program $S$ and observe a response~\cite{Eide:2008:VM:1450058.1450093}:
\begin{itemize}
\item Internal Compiler Error: $comp$ may crash during compilation because of a problem in $comp$ or $S$; for example a segmentation fault in GCC.
\item Functional Error: $comp(S)$ produces a compiled program $C$ that exhibits behaviour $B$ that differs from expected behaviour $B'$ when $C$ is run in a test environment $exec$; for example a run-time crash or concurrency bug.
\end{itemize}
We focus on concurrency bugs. A concurrency bug occurs when there are outcomes of executions of $comp(S)$ - run in test environment $exec$ - that are forbidden by $S$.

\begin{table*}[t]
\caption{Comparison of State-of-the-art Compiler Testing Techniques - inspired by Table 1 of ~\cite{DBLP:conf/asplos/BeckBSCBFV0H23}}\label{summary}
\centering
\begin{tabular}{|l|c|c|c|c|c|c|}
\hline
Technique & Automation & Coverage & General & Scalability & $exec$ &Comparison\\
\hline
Prose/Experts   & \xmark & ? & \cmark & \xmark & Human & Any\\
\texttt{cmmtest}~\cite{cmmtest} & ? & \xmark & \xmark & \xmark & Human+hardware & executions (def.~\ref{exe})  \\
\texttt{validc}~\cite{Chakraborty} & ? & \cmark & \xmark & \xmark & Human+models & executions \\
\texttt{C4}~\cite{10.1145/3460319.3469079,windsor2022high,9477685} & ? & \xmark & ? & \cmark & models+hardware & outcomes (def.~\ref{out}) \\
\telechat{} & \cmark & \cmark & \cmark & \cmark & models only & outcomes \\
\hline
\end{tabular}
\end{table*}

\begin{definition}{\textit{Concurrency Bug}}\label{fault}: for a multi-threaded $S$,

$outcomes_{C}(exec(C)) \not\subseteq outcomes_{S}(exec(S))$
\end{definition}

We test programs that exhibit bugs with at least two threads communicating via shared memory. Conversely, a \textit{negative difference} occurs when $outcomes_{C} \subset outcomes_{S}$ when optimisations are applied. The state-of-the-art make $exec$ precise (\textsection\ref{sota}). Like Leroy~\cite{Leroy:2009:FVR:1538788.1538814} we focus on deterministic programs (whose behaviour changes only in response to different initial states) and test environments (immune to changes in $exec$). We restrict bugs to executions that have different \textit{outcomes}. For instance, in Fig.~\ref{lb_exe}, the execution \textbf{dabc} - and its outcome \texttt{\{P1:r0=0; y=2\}} - is forbidden by the RC11 model~\cite{simonrc11} and ISO C/C++ model~\cite{C11}, but the compiled program allows it under the Armv8 model~\cite{aarch64}.

The choice of source model decides what is a bug. We use the RC11~\cite{Lahav:2017:RSC:3062341.3062352} model to explore the behaviours of Fig.~\ref{lb_rc11}, but emphasize that ISO C/C++ standard permits behaviours forbidden by RC11. Conversely, the Linux model~\cite{Alglave:2018:FSC:3173162.3177156} permits behaviours that are forbidden by standard C/C++~\cite{depslinux}. The source model thus acts as an oracle with respect to the allowed behaviours of the system under test. Since standards (and their models) can change - it is especially important to parameterise testing under multiple models. We support testing under models of source and compiled languages supported in mainstream C/C++ compilers. 

\begin{figure}[H]
\centering
  \begin{tabular}{c | c}
  Differential Testing & Metamorphic Testing\\
  \begin{tikzcd}[ampersand replacement=\&]
    \& S
      \arrow[dl, swap, bend right, "\texttt{comp$_1$}"]
      \arrow[dr, bend left, "\texttt{comp$_2$}"]
    \& \\
    C_1 
      \arrow[rr, dotted, "\text{bug?}"]
    \& 
    \& C_2 \\
  \end{tikzcd} &
  \begin{tikzcd}[ampersand replacement=\&]
    S \arrow[d, "\texttt{comp}"] \arrow[rr, "\texttt{variant}"]
    \& \& S_2 \arrow[d, "\texttt{comp}"] \\
    C_1 \arrow[rr, swap, dotted, "\text{bug?}"] \& \& C_2\\
  \end{tikzcd}
  \end{tabular}
  \caption{Differential and Metamorphic Testing - \textsection\ref{testing}.}
  \label{diff}
\end{figure}
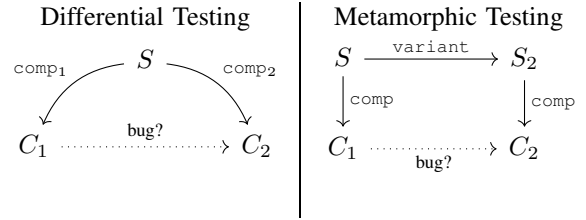

Chen et al.'s~\cite{10.1145/3363562} survey identifies two compiler testing techniques explored by the state-of-the-art: \textit{differential} and \textit{metamorphic} testing. Fig.~\ref{diff} illustrates both techniques. Differential testing (see CSmith~\cite{Yang}) compiles a program $S$ with different compilers, comparing the behaviours of each. For instance, comparing the outcomes of running executables produced by \texttt{clang -O1} and \texttt{clang -O3}. Metamorphic testing (see Orion~\cite{10.1145/2666356.2594334}) generates a variant $S_2$ of the source program $S_1$ that has the same behaviour as $S_1$, compiles both with the same compiler, and checks the behaviour of each is the same. For example, when compiling \texttt{print(1+1)} and \texttt{print(2)}, both compiled programs should output \texttt{2}.

We end this section mentioning related work that is out of scope. Donaldson et al. 2017~\cite{10.1145/3133917} and Lidbury et al. 2015~\cite{10.1145/2737924.2737986} test the compilation of GPU/OpenCL kernels, and graphics shaders. Neither test the compilation of concurrent programs~\cite{Chakraborty} in the C/C++ sense, as multi-threaded GPUs support a parallel computation model.

\subsection{State-of-the-art Techniques}\label{sota}

We summarise the state-of-the-art in compiler testing with memory models. We compare works in terms of \textit{Automation}, \textit{Coverage}, \textit{Generality}, and \textit{Scalability} (see \textsection\ref{sec:intro}). A tool is \textit{automatic} if it can be used by compiler engineers in regression testing with no intervention by concurrency experts to generate tests or interpret results. \textit{Coverage} describes the set of potential bugs the tool will discover. \textit{Scale} bounds the number of threads and LoCs of inputs. A tool is \textit{general} if it supports multiple source and compiled languages. In Table~\ref{summary} we state whether the solution fulfils the requirement with a \cmark, does not \xmark, and partially fulfils the requirements with $?$.

\subsubsection{Prose and Expertise} The first (non-)solutions involve reading prose language standards such as ISO C/C++~\cite{C11}, or consulting memory model experts. Both approaches are manual and are effective in finding bugs. Both are prohibitively tedious and expensive for use in routine regression testing.

\subsubsection{Semi-automatic Tools} The \texttt{cmmtest} tool~\cite{cmmtest} conducts differential testing by extending CSmith with concurrency support for an early C/C++ model. Given a \textit{single-threaded} C/C++ program, \texttt{cmmtest} checks if the hardware execution of the optimised program is a sub-graph of an unoptimised hardware execution, else a bug may occur and a concurrency expert finds a test case reproducer. \texttt{validc}~\cite{Chakraborty} builds on \texttt{cmmtest}~\cite{cmmtest} by matching \textit{all} bounded executions of optimised LLVM IR programs against unoptimised IR.

Both techniques are manual as experts must reproduce bugs using the warnings output by the tools. Since execution matching is an instance of the sub-graph isomorphism problem~\cite{10.1145/800157.805047} it will not scale in general~\cite{DBLP:phd/ethos/Nimal14}. Neither technique is general, as \texttt{cmmtest} relies on x86-only~\cite{pin} tools, and \texttt{validc} accepts only LLVM IR programs. The \texttt{validc} tool covers bugs allowed by the C/C++ or LLVM models; however \texttt{cmmtest} may miss bugs as it relies on hardware.

\subsubsection{Hardware-based Tools} The \texttt{C4} tool of Windsor et al. 2021/22~\cite{10.1145/3460319.3469079,windsor2022high,9477685} conducts metamorphic testing of litmus tests by comparing the outcomes of hardware runs (using the \texttt{litmus} tool) against outcomes of source test simulations (using the \texttt{herd} tool) under the RC11 model~\cite{simonrc11}. \texttt{C4} is automatic and testing scales as hardware often runs quickly. Hardware test environments are nondeterministic and may omit behaviour - and hence bugs - allowed by an architecture model (\textsection\ref{tools}). To improve coverage, Windsor et al. ``stress-test''~\cite{10.1145/3460319.3469079} hardware. Since \texttt{C4} was developed in parallel with our work, we summarise it:
\begin{multline*}\label{eq:c4}\tag{$\textnormal{\textbf{test}}_{C4}$}
outcomes(\texttt{litmus}(comp(S),hardware)) \\
  \subseteq outcomes(\texttt{herd}(S,M_{S}))
\end{multline*}

\subsubsection{Our Solution - Testing with Models}

The \texttt{herd} tool~\cite{Alglave:2014:HCM:2633904.2627752} simulates \textit{both} source and compiled litmus tests under source $M_S$ and architecture models $M_C$ automatically. It follows that we can test compilers by comparing the outcomes (see Fig.~\ref{lb_log}) of executions of compiled programs under $M_C$ against outcomes executions of a source program under $M_S$.

\section{Design and Implementation of \telechat{}}\label{sec:wip}
\begin{figure}[H]
\centering
  \begin{tikzcd}[ampersand replacement=\&, row sep=large, column sep=large]
    entry \arrow[r, "\text{1. generate (\texttt{diy})*}"]
    \& S
      \arrow[r, "\text{2. \telechat{}}"]
      \arrow[d, swap, "3. \text{\texttt{herd}($S$, $M_{S}$)*}"]
    \& C
      \arrow[d, "\text{4. \texttt{herd}(C, $M_{C}$)*}"]\\
    \& outcomes_S
      \arrow[r, leftarrow, dotted, "5.\supseteq  (\texttt{mcompare})*"]
    \& outcomes_C
    \end{tikzcd}
  \caption{Test environment $exec_{tv}$ including the \telechat{} tool and tools we improved marked *. \ref{eq:tv} checks the outcomes of simulating $comp(S)$ under its architecture model $M_C$ against the source $S$ under source model $M_S$.}\label{tele}
\end{figure}
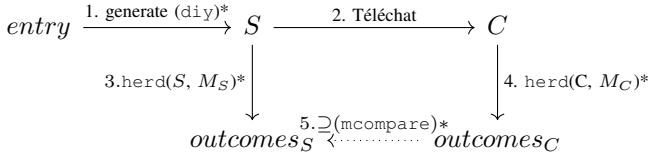

\subsection{Technique Design}\label{technique}
We present the \telechat{} automatic testing tool and technique \ref{eq:tv}; for compilers including GCC and LLVM. Fig.~\ref{tele} details the test environment $exec_{tv}$ summarised by:
\begin{multline*}\label{eq:tv}\tag{$\textnormal{\textbf{test}}_{tv}$}
outcomes(\texttt{herd}(comp(S),M_{C})) \\
  \subseteq outcomes(\texttt{herd}(S,M_{S}))
\end{multline*}
The test environment of Fig.~\ref{tele} ($exec_{tv}$) proceeds as follows:

\begin{enumerate}
\item Generate concurrent C/C++ litmus test $S$.
\item \telechat{} prepares $S$ for compilation, compiles it using $comp$ and disassembles the relocatable ELF file, then constructs an assembly litmus test $C$ and state mappings $m$ from outcomes of $S$ to outcomes of $C$.
\item Simulate $S$ using \texttt{herd} under one of the C/C++ memory models in the herd tool-suite~\cite{herdtools} (see \textsection\ref{tools}), collect allowed C/C++ outcomes $outcomes_S$.
\item Simulate $C$ using \texttt{herd} under its architecture memory model in the herd tool-suite~\cite{herdtools} (see \textsection\ref{tools}). Get architecturally allowed outcomes $outcomes_C$.
\item Use \texttt{mcompare} from the herd tool-suite~\cite{herdtools} to check if $outcomes_C \subseteq outcomes_S$ using state mappings $m$. If $outcomes_C \not\subseteq outcomes_S$ then there is a bug.
\end{enumerate}

\telechat{} enables the automatic testing of program outcomes by completing the graph in Fig.~\ref{tele}. To support compiled tests $comp(S)$, we extend the \texttt{diy}~\cite{Alglave:2012:FWM:2205506.2205523} test generator, the \texttt{herd}~\cite{Alglave:2014:HCM:2633904.2627752} simulator, and \texttt{mcompare}~\cite{herdtools}.

The \ref{eq:tv} technique is remarkably simple. By checking that the \textit{expected} outcomes of a source test under the source memory model include the \textit{actual} outcomes of the compiled test under an architecture model we get a technique that is familiar to engineers who are not necessarily experts in relaxed memory concurrency. \ref{eq:tv} is simple enough that \telechat{} is cited in discussions by Arm's engineers~\cite{LDAPR}.

$exec_{tv}$ is a deterministic (\textsection\ref{testing}) test environment. In other words, we compare outcomes under source and architecture models - rather than relying on hardware or the operating system. Further, \texttt{herd} runs deterministic litmus tests: from a fixed initial state with a fixed loop unroll factor, under models of Armv8 AArch64~\cite{aarch64} (official), RISC-V~\cite{lucrisc} (official), RC11~\cite{simonrc11}, Armv7 (unofficial)~\cite{aarch32}, Intel x86-64~\cite{x86tso}, MIPS~\cite{mips}, IBM PowerPC~\cite{ppc}, and more.

The \telechat{} tool-chain is run as part of regular automated compiler testing and is, as far as we know, the industry's first publicly available compiler testing tool that is deployed in automated compiler testing of concurrent C/C++. Nothing prevents deployment of \telechat{} outside of Arm however.

\subsection{Tool Implementation} \label{im}
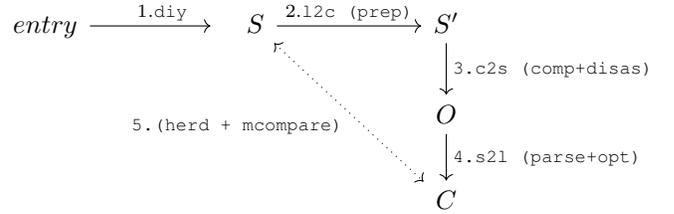
\begin{figure}[h]
  \centering
\begin{tikzcd}[ampersand replacement=\&, column sep=5.5em]
  entry \arrow[r, shorten >=2ex, "1.\texttt{diy}"]
  \& S \arrow[r, "2. \texttt{l2c (prep)}"] \arrow[ddr, leftrightarrow, dotted, "\texttt{5.(herd + mcompare)}" swap]
  \& S' \arrow[d, "\texttt{3.c2s (comp+disas)}"]\\ 
  \& \& O \arrow[d, "\texttt{4.s2l (parse+opt)}"] \\
  \& \& C
\end{tikzcd}
  \caption{Breakdown of the \telechat{} tool. The \texttt{l2c}, \texttt{c2s}, and \texttt{s2l} tools are new. We modified \texttt{diy}~\cite{Alglave:2012:FWM:2205506.2205523}, \texttt{herd}~\cite{Alglave:2014:HCM:2633904.2627752}, and \texttt{mcompare}~\cite{herdtools} to to accept compiler-generated tests.}
  \label{impl}
\end{figure}

Fig.~\ref{impl} breaks down Step 2 (\telechat{}) of Fig.~\ref{tele} as follows:
\begin{enumerate}
\item \textit{Input}: Generate a C/C++ litmus test $S$ (step 1 of Fig.~\ref{tele}).
\item The \textit{litmus2c} (\texttt{l2c}) tool prepares $S$ for compilation, producing a C/C++ program $S'$. Optionally fuzz $S'$.
\item The \textit{c2assembly} (\texttt{c2s}) tool compiles $S'$ with ELF relocations (requires flags \texttt{-c -g}) and disassembles the object file using GNU or LLVM \texttt{objdump}. \texttt{c2s} returns an assembly file $O$ and state mappings $m$.
\item The \textit{assembly2litmus} (\texttt{s2l}) tool parses $O$ and constructs an optimised assembly litmus test $C$.
\item \textit{Output}: Pass $S$ and $C$ to \texttt{herd} for simulation under source and architecture models (steps 3-4 of Fig.~\ref{tele}).
\end{enumerate}

\subsection{Adoption, Usage, and Documentation}\label{adopt}

We provide a Docker artefact to reproduce our work. The artefact contains builds of LLVM and GCC for the architectures above, the \telechat{} tool, and a Makefile to reproduce the experiments. A variant of the Docker artefact is deployed in automated testing for Arm's compiler teams.

For users we provide guides and example test suites. \telechat{} ships a user guide with its artefact. The Docker file contains example tests suites to build on.

To reduce the risk of tool stagnation, we provide Arm's compiler team with documentation including internal conference talks on \telechat{}, memory model training, and wiki pages. Arm's compiler engineers are increasingly using litmus tests to discuss concurrency queries as they arise~\cite{john}. 

\subsection{Challenges Encountered During Implementation}

We end by discussing the challenges encountered whilst implementing \telechat{}. The first challenge arose in how to represent compiled programs as litmus tests. Compiled programs represent memory locations as binary addresses \texttt{0xf00} that can be manipulated with arithmetic instructions. ELF files layout multiple locations together in sections. Litmus tests represent memory locations as symbolic variables \textit{x} that have no memory layout constraints. We use DWARF metadata to map numeric addresses to symbolic locations and symbol table information to gather memory layout constraints. As such our technique is as accurate as the metadata compilers provide. By converting between address formats, \telechat{} bridges a gap between formal modelling tools and real-world systems.

Developing \telechat{} spawned many extensions to the herd tool-suite~\cite{herdtools}, including formalising the semantics of new instructions, new data types, and tools. For instance we added a vector datatype to model memory layout and store pair instructions that span contiguous locations. We also added a regression suite for the herd tool-suite itself. Our work was used by students in projects modelling the NEON~\cite{neon} and SVE~\cite{sve} extensions of the Arm architecture. To compare outcomes of tests of differing architectures or languages, we added state mapping support to \texttt{mcompare}. 

\section{Evaluation}\label{sec:eval}
We evaluated \telechat{} by conducting experiments using multiple compilers. Our results suggest \telechat{} improves on the state-of-the-art (we found one existing behaviour missed by \texttt{C4}, and four new bugs~\cite{ldpbug,constbug,stp,swp}), proposes new lines of inquiry into test generation (a new kind of bug), questions interactions between sequential and concurrency semantics (\texttt{const} and \texttt{atomic}), and makes an impact in industry. 

\subsection{Comparison with The State-Of-The-Art: C4}\label{c4}

\begin{table*}[t]
  \centering
  \caption{C4 versus \telechat{}.} 
  \begin{tabular}{| l | c | c |}
  \hline
  Component & \texttt{C4}~\cite{10.1145/3460319.3469079,windsor2022high,9477685}& \telechat{} \\
  \hline
  Test Generator - \textsection\ref{tools} & \texttt{Memalloy}~\cite{Wickerson:2017:ACM:3009837.3009838} & \texttt{diy}~\cite{Alglave:2012:FWM:2205506.2205523}\\
  Test Environment - \textsection\ref{testing} & models+hardware & models only  \\
  Source $exec$ - \textsection\ref{testing} &\texttt{herd}~\cite{Alglave:2014:HCM:2633904.2627752} & \texttt{herd}\\
  Target $exec$ &\texttt{litmus}~\cite{Alglave:2011:LRT:1987389.1987395} & \texttt{herd}\\
  Testing method - \textsection~\ref{testing} & Metamorphic Testing~\cite{10.1145/2666356.2594334} & Metamorphic \& Differential Testing~\cite{Yang} \\
  Models involved & source & source and architecture\\
  System under test (SUT) & Compiler + Hardware + OS & Compiler \\
  Found Bugs? & Yes & Yes~\cite{ldpbug,constbug,stp,swp} \\
  Automatic & No (must stress SUT) & Yes\\
  Coverage & No & Yes (up to fixed bounds) \\
  General & No & Yes (parameterised over models)\\
  Scalable & Yes & Yes\\
  Deterministic & No & Yes \\
  \hline
  \end{tabular}
  \label{compare}
\end{table*}

\telechat{} is similar to \texttt{C4}~\cite{10.1145/3460319.3469079,windsor2022high,9477685}, but \telechat{} relies only on simulation. Both tools rely on the \texttt{herd} tool-suite~\cite{herdtools}. \telechat{} differs in that it relies \textit{solely} on simulation of both source and compiled litmus tests whereas \texttt{C4} relies on hardware executions to collect compiled test outcomes. Tab.~\ref{compare} summarises the differences between \texttt{C4} and \telechat{}. This small change in technique has consequences for whether behaviours are observable in the test environment. 

\begin{figure}[t]
\centering
  \begin{tabular}{c}
\begin{lstlisting}[language=C, style=mystyle]
{ *x = 0, *y = 0 }

#define relaxed memory_order_relaxed
#define load atomic_load_explicit
#define store atomic_store_explicit

void P0(atomic_int* y,atomic_int* x) {
  int r0 = load(x,relaxed);
  atomic_thread_fence(relaxed);
  store(y,1,relaxed);
}
void P1(atomic_int* y,atomic_int* x) {
  int r0 = load(y,relaxed);
  atomic_thread_fence(relaxed);
  store(x,1,relaxed);
}

exists (P0:r0=1 /\ P1:r0=1)
\end{lstlisting}
  \end{tabular}
  \caption{The outcome \texttt{\{ P0:r0=1; P1:r0 = 1\}} is forbidden under the proposed RC11~\cite{Lahav:2017:RSC:3062341.3062352} model. When compiled the outcome is allowed by Armv8 AArch64~\cite{aarch64}, Armv7~\cite{aarch32}, PowerPC~\cite{ppc}, and RISC-V~\cite{lucrisc} models.}
  \label{lb_c4}
\end{figure}

We compare tools directly as both take litmus tests and compare outcomes. We pass 85 litmus tests to both tools and compare the outcomes (def.~\ref{out}) of each. \telechat{} finds behaviours that Windsor et. al miss~\cite{windsor2022high}. Consider Fig.~\ref{lb_c4} and its outcomes when simulated under the RC11 model~\cite{simonrc11} - Fig.~\ref{lb_aarch64} (left). We compile Fig.~\ref{lb_c4} using \telechat{} and LLVM-11 (\texttt{clang -O3}) to get an Arm AArch64 litmus test. Fig.~\ref{lb_aarch64} (right) shows the outcomes of simulating the assembly litmus test under the Arm AArch64 model~\cite{aarch64}. We observe the outcome \texttt{\{ P0:r0=1; P1:r0 = 1\}} that is forbidden by the RC11 model (Fig.~\ref{lb_aarch64} (left)), but allowed by the AArch64 model (Fig.~\ref{lb_aarch64} (right)). Windsor et. al state that \texttt{C4} missed this behaviour, but they observe it under model simulation, which increases our confidence in \telechat{}.

\begin{figure}[h]
  \begin{tabular}{c || c}
  RC11~\cite{Lahav:2017:RSC:3062341.3062352} Outcomes & Arm AArch64 Outcomes \\
\begin{lstlisting}[language=C, style=mystyle]
{P0:r0=0; P1:r0=0;}
{P0:r0=0; P1:r0=1;}
{P0:r0=1; P1:r0=0;}
.
\end{lstlisting} &
\begin{lstlisting}[language=C, style=mystyle]
{P0:r0=0;P1:r0=0;}
{P0:r0=0;P1:r0=1;}
{P0:r0=1;P1:r0=0;}
{P0:r0=1;P1:r0=1;}<-C4 missed
\end{lstlisting}\\
  \end{tabular}
  \caption{(left) Fig.~\ref{lb_c4} outcomes allowed by the RC11 model~\cite{simonrc11}. (right) Outcomes of \telechat{}-generated test allowed by the Arm AArch64 model~\cite{aarch64}.}
  \label{lb_aarch64}
\end{figure}

We found hundreds of litmus tests that induce this behaviour under RC11~\cite{simonrc11} when compiled by either LLVM or GCC, detailed in \textsection\ref{large}. Fig.~\ref{lb_c4} implements the \textit{load buffering} (LB) pattern, known by concurrency experts.  Further, we observe the same behaviour when compiling to target Armv7 (unofficial), IBM PowerPC, and RISC-V (official).

We conclude that \telechat{} is deterministic unlike \texttt{C4}. In other words, \telechat{} observes the same test outcomes every time. \texttt{C4} requires that the hardware exhibits an outcome and that users `stress-test'~\cite{10.1145/3460319.3469079} the hardware to reproduce it. Silicon manufacturers may however implement restricted variants of an architecture model (\textsection\ref{tools}). \texttt{C4} is not guaranteed to observe the same outcomes on different machines, or even the same machine. Indeed, Sarkar et. al~\cite{Sarkar:2011:UPM:1993498.1993520} observe LB on an Apple A9 and Nvidia Tegra2 chips\footnote{\url{https://www.cl.cam.ac.uk/~pes20/arm-supplemental/arm001.html\#toc5}}, but Windsor et. al miss it~\cite{windsor2022high} using a Raspberry Pi. It is possible that developing \texttt{C4}'s metamorphic relations may increase the chance of finding bugs, provided the hardware provides a witness to miscompilation.

\telechat{} is useful when hardware is inaccessible. For instance, we assisted Arm's engineers with a query from an Arm partner, who proposed to change the compilation of C/C++ atomic acquire loads when targeting Armv8.3-a~\cite{LDAPR}. The proposed change had promising performance characteristics on unspecified hardware, but correctness was untested beyond interpreting the Armv8.3-a specification. Arm's compiler teams accepted the proposal based on our findings.

\telechat{} does not completely subsume the state-of-the-art as, for example, simulation does not terminate when checking huge systems (with thousands of LoC). \texttt{C4} can do this. We expect the success of \telechat{} depends on validity of the \textit{small-scope hypothesis}, which we explore in \textsection\ref{limit}.

\subsection{The Local Variable Problem} \label{local}

The local variable problem affects all state-of-the-art techniques (\textsection\ref{sota}) and masks a kind of bug that has evaded detection. We discovered that, contrary to the state-of-the-art, there are optimisations affecting only thread-local state that influence concurrent program execution, giving rise to a new class of bug. The problem is that there are transformations allowed by the C/C++ model~\cite{Chakraborty} that delete data required to detect bugs. We reported a new bug of this kind~\cite{swp} (see Fig.~\ref{lb_rc11}), reproduced two bugs for Arm's engineers, and present our solution in \telechat{}.

Consider the load buffering (LB) litmus test in Fig.~\ref{lb_local}. When simulated using \texttt{herd}~\cite{Alglave:2014:HCM:2633904.2627752} the values of the local variables \texttt{P0:r0} and \texttt{P1:r0} are recorded for use when checking outcomes. C/C++ memory models~\cite{simonrc11} allow the compiler to delete unused local data. Consequently, a litmus test that refers to \textit{deletable} data~\cite{Chakraborty} in its final state - like \texttt{P0:r0} and \texttt{P1:r0} - will have no data to refer to if the compiler removes it. When compiling Fig.~\ref{lb_local} (left) with \texttt{clang -O2} we get Fig.~\ref{lb_local} (right -  in C for illustration purposes). The only allowed outcome of Fig.~\ref{lb_local} (right) is \texttt{\{ P0:r0=0; P1:r0=0 \}} since \texttt{herd} assumes data is zero-initialised.

\begin{figure}[h]
\centering
  \begin{tabular}{c | c}
\begin{lstlisting}[language=C, style=mystyle]
{ *x = 0, *y = 0 }

void P0 (int* y,int* x) {
  int r0 = *x; // unused
  *y = 1;
}
void P1 (int* y,int* x) {
  int r0 = *y; // unused
  *x = 1;
}
exists
  (P0:r0=1 /\ P1:r0=1)
\end{lstlisting} &
\begin{lstlisting}[language=C, style=mystyle]
{ *x = 0, *y = 0 }

void P0 (int* y,int* x) {
  // deleted
  *y = 1;
}
void P1 (int* y,int* x) {
  // deleted
  *x = 1;
}
exists
  (P0:r0=1 /\ P1:r0=1)
\end{lstlisting}
  \end{tabular}
  \caption{(left) Load Buffering (LB) litmus test. (right) Load Buffering test after \texttt{clang -O2} deletes unused data.}
  \label{lb_local}
\end{figure}

Local reads of shared data are snapshots of that data at particular points in a  program's execution. Several concurrency patterns rely on local data to convey whether the reordering of accesses leads to forbidden outcomes. For example, LB demonstrates a notion of locality - for instance caches - that \textit{buffer} loads during execution. Local data reordering is common in many processors - we cannot ignore it, even if the C/C++ model permits its removal.

Testing techniques may miss bugs in optimisations that delete local data. Such techniques cannot test the compilation of LB unless local data persists. It is possible that an assembly language register of the compiled program contains local data, but compilers often reuse registers to reduce spilling. The state-of-the-art (\textsection\ref{sota}) relies \textit{only} on source (e.g C/C++) models - without testing if local deletion masks bugs. 

Authors either overlook the issue~\cite{Chakraborty,10.1145/3460319.3469079} or claim~\cite{cmmtest} local optimisations cannot induce bugs (def.~\ref{fault}). When asked about local optimisations at the European LLVM conference (2017)\footnote{33:50 minutes in: \url{https://www.youtube.com/watch?v=NR5OAhgdozc}}, Chakraborty and Vafeiadis~\cite{Chakraborty} state they focus on ``only the shared memory accesses''. Windsor et. al do not address the issue~\cite{10.1145/3460319.3469079}. Morisset et al.~\cite{cmmtest} claim that ``optimisations affecting only the thread-local state cannot induce concurrency compiler bugs''. We question this~\cite{cmmtest} claim.

\begin{figure}[h]
\centering
\begin{tabular}{c}
\begin{lstlisting}[language=C, style=mystyle]
{ *x = 0, *y = 0 }

#define relaxed memory_order_relaxed

void P0 (atomic_int* y,atomic_int* x) {
  atomic_store_explicit(x,1,relaxed);
  atomic_thread_fence(memory_order_release);
  atomic_store_explicit(y,1,relaxed);
}
void P1 (atomic_int* y,atomic_int* x) {
  int r1 = atomic_fetch_add_explicit(y,1,relaxed);
  atomic_thread_fence(memory_order_acquire);
  int r0 = atomic_load_explicit(x,relaxed);
}
exists (P1:r0=0 /\ y=2) 
\end{lstlisting}
\end{tabular}
\setlength{\textfloatsep}{2pt}
\caption{Message Passing litmus test. The outcome \texttt{\{P1:r0=0; y=2\}} is forbidden under the C/C++ model~\cite{simonrc11}, but allowed by the Arm AArch64 model~\cite{aarch64}, since an LLVM thread-local optimisations can remove \texttt{P1:r1}. The heisenbug arises if \texttt{P1} does not observe the read of the RMW operation.}
\label{fig:counter}
\end{figure}

Fig.~\ref{fig:counter} induces a bug when thread-local optimisations delete \texttt{P1:r1}. The outcome \texttt{\{P1:r0=0; y=2\}} is forbidden by the C/C++ model, but allowed by the LLVM or GCC compilation to Arm AArch64 when the assignment to \texttt{P1:r1} by the read-modify-write (RMW) operation is deleted. Past versions of LLVM and GCC induce this bug when targeting Armv8.1-a with the Large-Systems Extension. This example shows that thread-local optimisations can induce concurrency bugs.

\texttt{P1} uses an atomic \texttt{fetch\_add\_explicit} RMW operation where the value read into \texttt{P1:r1} is unused. This induces \textit{two} bugs using past versions of LLVM and GCC, first by targeting the incorrect Arm instruction and second by deleting \texttt{P1:r1}. In both cases the outcome \texttt{\{P1:r0=0; y=2\}} is allowed by the LLVM or GCC compilation to Arm AArch64, but forbidden by the C/C++ model. Engineers fixed the first bug replacing the \texttt{STADD} instruction with \texttt{LDADD}. The second bug is observed when the LLVM dead register definitions pass~\cite{LLVMdrd} zeroes the destination-register of \texttt{LDADD}, after \texttt{P1:r1} is deleted - the bug is observed as \texttt{LDADD} aliases \texttt{STADD} when the destination register is the zero register.

Finding these bugs without \telechat{} required expertise and two engineer years of work. Discussions involved Linux Kernel maintainers, Arm AArch64 model authors, and GCC (and LLVM) developers~\cite{llvm-stadd-2,llvm-stadd-1,gcc-stadd-1,gcc-stadd-2}. We assisted Arm's compiler teams by reproducing the bugs and showing that the latest versions of LLVM and GCC no longer exhibit them. We added support for Fig.~\ref{fig:counter} (with and without \texttt{int r1 = ...}) to \texttt{herd}~\cite{dmbish}, allowing us to validate the fix. We reported~\cite{swp} a new bug of this kind in the implementation of \texttt{atomic\_exchange} featured in Fig.~\ref{lb_rc11}, but it is unclear if more bugs like this exist.

Interestingly, these bugs disappear if one attempts to study them. Historically, message passing tests check the reordering of \texttt{P1:r0} and \texttt{P1:r1} - forcing the user to preserve local data. If instead we delete \texttt{P1:r1} and check the value of \texttt{y} in the final state, then we see reordering. In other words, you only find the bug through \textit{indirect} observation - it is a new kind of Heisenbug! Since current test generators implement the historical case, it is no surprise that these bugs were discovered manually until now.

We implement a solution in \telechat{}. \telechat{} augments Fig.~\ref{lb_local} with global variables that store local data at the end of each thread. This augmentation is optional to allow thread-local optimisations to be tested. The original code under test remains, but with the additional constraint that local data persists after compilation. We update the initial and final states to reflect this new data. It is unsatisfactory to modify the test, but we have found four bugs thus far~\cite{ldpbug,constbug,stp,swp}. We are open to better solutions, if they exist.

\subsection{Bug-Finding Campaign}\label{bugs}
Whilst developing \telechat{} we reported two new concurrency bugs~\cite{ldpbug,stp} in LLVM, and a missed optimisation~\cite{mipbug} for the MIPS backend of GCC. We propose two bug fixes that Arm's engineers are addressing, and note one line of inquiry for compiling atomics in practice -  all untested until now.

First, we reported a bug~\cite{ldpbug} in the compilation of 128-bit sequentially consistent~\cite{Lamport:1979:MMC:1311099.1311750} loads. The bug occurs when a sequentially consistent~\cite{Lamport:1979:MMC:1311099.1311750} atomic load is implemented using a load pair instruction on Armv8.4. The Armv8.4 Large Systems Extension (v2) ensures load or store pair instructions are single-copy atomic~\cite{Seal:2000:AAR:517257}, assuming accesses are 16-byte aligned to normal memory. This means you can use an \texttt{LDP} instruction in place of a potentially more expensive \textit{compare-and-swap} (CAS) loop. \texttt{LDP} has no ordering requirements however - \texttt{LDP} can be reordered before a prior store of an atomic read-modify-write operation that uses a CAS loop. We propose to fix sequential consistency in LLVM by adding synchronisation, following GCC~\cite{GCCldpfix}.

Next, we reported a wrong-endian bug~\cite{stp} in the compilation of 128-bit atomic stores. Since AArch64 has 64-bit register sizes, an 128-bit store is implemented using a \textit{pair} of 64-bit registers. We report that the order registers are written to memory is flipped by atomic store operations. This affects store-release-exclusive pair instructions in CAS loops (for Armv8.3 or below), and individual store pair instructions (Armv8.4 or above). We propose to flip the bits to fix the bug.

We reported~\cite{mipbug} an optimisation opportunity in the MIPS backend of GCC. Whilst developing \telechat{}, we discovered that GCC (and LLVM) are conservative in optimising instructions that access \texttt{atomic} data. Extra code is emitted, since \texttt{atomic}-accessing code cannot inhabit branch delay slots. GCC maintainers note that \texttt{atomic} data is considered \texttt{volatile} for practical reasons, despite no change in compiled program outcomes (def.~\ref{out}) under models. Whether it is \textit{still} valid to treat \texttt{atomic} as \texttt{volatile} is further work. The above bugs are in dark corners, difficult for experts to find even when an excellent memory model exists. Without \telechat-style automation these faults are impossibly expensive to test for in routine regression testing.

\subsection{Large-Scale Differential Testing}\label{large}

\begin{table*}[t]
\centering
\caption{We test combinations of $\textnormal{C/C++ constructs} \times \textnormal{Compiler Under Test} \times \textnormal{Flags} \times \textnormal{Arch}$.}
\begin{tabular}{l | c}
C/C++ constructs: & \texttt{(atomic operations|non-atomic operations}\\
& \texttt{ |fences|control-flow|straight-line code)+}\\
Compiler under test: & \texttt{(LLVM|GCC)}\\
Optimisation flags: & \texttt{(-O1|-O2|-O3|-Ofast|-Og)+}\\
Target Architecture: &\texttt{(Armv8 AArch64 (64-bit official)|Armv7-a (32-bit unofficial)|RISC-V (64-bit official)}\\
& \texttt{|Intel x86-64 (64-bit)| MIPS (64-bit)|IBM PowerPC (64-bit))}
\end{tabular}
\label{fig:wow}
\end{table*}

\begin{table*}[t]
\caption{Test Results - Takes 9 hours and 40 minutes on a 224 Core ThunderX2 using 100GB runtime footprint. clang does not support \texttt{-Og} flag. 167,184 C tests input, 9,027,936 compiled tests output, total: 9,195,120. Total \% = sum(row)/compiled tests output * 100 (3 sf). These results were collected using the RC11 model~\cite{Lahav:2017:RSC:3062341.3062352}, all positive differences disappear if load buffering is permitted.}
\begin{center}

  \begin{tabular}{ |l||c c c c c | l | } 
  \hline
  & {\tt -O1} & {\tt -O2} & {\tt -O3} & {\tt -Ofast} & {\tt -Og} & Total \% \\
  \hline
  Armv8 AArch64 (64-bit) & \multicolumn{5}{c|}{clang/gcc} & \\
  +ve & 2352/2352 & 2352/2352 & 2352/2352 & 2353/2352 & -/2352 & 0.23\% \\ 
  -ve & 44300/44300 & 44300/44300 & 44300/44300 & 44300/44300 & -/44300 & 4.42\%\\
  \hline
  Armv7-a (32-bit) & \multicolumn{5}{c|}{clang/gcc} &\\
  +ve & 2352/3480 & 2352/2352 & 2352/2352 & 2352/2352 & -/2352 & 0.25\% \\ 
  -ve & 68228/69890 & 68228/70220 & 68228/70220 & 68228/70220 & -/70220 & 6.91\%\\
  \hline
  RISC-V (64-bit) & \multicolumn{5}{c|}{clang/gcc} & \\
  +ve & 2352/2352 & 2352/2352 & 2352/2352 & 2352/2352 & -/2352 & 0.23\% \\ 
  -ve & 34204/70772 & 34204/70772 & 34204/70772 & 34204/70772 & -/70772 & 5.44\% \\
  \hline
  IBM PowerPC (64-bit) & \multicolumn{5}{c|}{clang/gcc} & \\
  +ve & 2352/2352 & 2352/2352 & 2352/2352 & 2352/2352 & -/2352 & 0.23\% \\ 
  -ve & 43956/43956 & 43956/43956 & 43956/43956 & 43956/43956 & -/43956 & 4.38\%  \\
  \hline
  Intel x86-64 (64-bit) & \multicolumn{5}{c|}{clang/gcc} & \\
  +ve & 0/0 & 0/0 & 0/0 & 0/0 & -/0 & 0.0\%\\ 
  -ve & 64112/64112 & 64112/64112 & 64112/64112 & 64112/64112 & -/64112 & 6.39\%\\
    \hline
  MIPS (64-bit) & \multicolumn{5}{c|}{clang/gcc} & \\
  +ve & 0/0 & 0/0 & 0/0 & 0/0 & -/0 & 0.0\%\\ 
  -ve & 69664/72488 & 69664/72008 & 69664/72008 & 69664/72008 & -/72488 & 7.09\% \\
  \hline
  \end{tabular}
\end{center}
\label{fig:results}
\end{table*}

We use \telechat{} to conduct differential testing of LLVM and GCC. We check compatibility between compilers, as code generated by LLVM and GCC is often mixed together at link-time or by operating systems, potentially inducing latent bugs at runtime. We test compilation targeting multiple architectures using commonly used flags for a large suite of tests. We ran over 9 million tests that have 2 to 5 threads, up to 5 shared variables, and up to 50 lines of compiled assembly code. On each thread Téléchat removes around 4 lines of (compiled) code per access. As far as we know this is the most extensive concurrency testing campaign to date. We test:

\begin{itemize}
\item Compilers: LLVM and GCC compiling C/C++ to target Armv8 AArch64 (64-bit), Armv7-a (32-bit), RISC-V, Intel x86-64, IBM PowerPC, and MIPS.
\item Optimisation levels for each compiler: \texttt{-O1}, \texttt{-O2}, \texttt{-O3}, \texttt{-Ofast}, and for GCC \texttt{-Og}.
\item Compare a compiler with itself at increasing levels of optimisation, e.g. \texttt{clang -O1} vs. \texttt{clang -O2}.
\item Compare LLVM with GCC at each optimisation level, e.g \texttt{clang -O1} vs. \texttt{gcc -O1}.
\end{itemize} 

Tab.~\ref{fig:wow} defines all the combinations of test, compiler, and architecture under test. Our tests feature code that perturbs the order accesses hit memory including control-flow, atomic operations, non-atomic operations, fences, and straight-line code. We test using both signed and unsigned integers ranging from 8-bits up to 64-bits in size. We test both LLVM and GCC with the optimisations and architectures above.

Following the steps in \textsection\ref{technique}, we generate multiple source C/C++ \textit{test sets} enumerating the features in Tab.~\ref{fig:wow} using \texttt{diy}~\cite{Alglave:2012:FWM:2205506.2205523}. For each test set, we use \telechat{} with the compiler under test to generate {\it multiple} assembly test sets according to multiple {\it compiler profiles}. Each profile captures the compiler tool-chain (\& flags), architecture (\& model), disassembler (\& flags), and symbol table reader. For instance, the \texttt{llvm-O3-AArch64} profile tests: \texttt{clang -O3} using the AArch64 GNU/Linux bare-metal tool-chain and \texttt{gnu-objdump}. Both source and target tests are passed to \texttt{herd} for simulation under the RC11~\cite{simonrc11} and Arm AArch64 model~\cite{aarch64} respectively (resp. target architecture models). Lastly, \texttt{mcompare} compares outcomes to find outcomes (def.~\ref{out}) of the compiled program $outcomes_C$ that are not outcomes of the source program $outcomes_S$:

\begin{itemize}
\item \textit{positive differences (+ve)}: $outcomes_C \not\subseteq outcomes_S$
\item \textit{negative differences (-ve)}: $outcomes_C \subset outcomes_S$.
\end{itemize}

(negative differences can occur since both optimisations and architecture models can constrain behaviour).

Tab.~\ref{fig:results} details our results. Tab.~\ref{fig:results} suggests \telechat{} is effective as it found tricky concurrency behaviours hidden in over 9 million compiled tests, given 167,184 tests as input. The 2352 positive differences common to Armv8 (official), Armv7 (unofficial), RISC-V (official), and IBM PowerPC are due to 294 variants of the load buffering pattern in Fig~\ref{lb_c4}. When comparing \texttt{llvm-O1-ARM} +ve (2352) and \texttt{gcc-O1-ARM} +ve (3480) we discovered two behaviours in GCC and LLVM. Re-ordering is observed using \texttt{-O1} when a control dependency is removed, but the behaviour is masked at higher optimisation levels by a data dependency (\texttt{-O2} and above). Since Intel x86-64 implements the total-store order model~\cite{Sarkar:2009:SXM:1480881.1480929} there are are no differences. This suggests \telechat{} is an effective compiler testing tool.

To be clear, these positive differences are not \textit{bugs} in today's compilers, since we used the RC11 model~\cite{Lahav:2017:RSC:3062341.3062352} that is not ratified by the C/C++ standards. The ISO C/C++ standards explicitly permit load-to-store reordering (\textsection7.17.3 of C23~\cite{C11}), whereas RC11 forbids it. \telechat{} is parameterised over models, and we repeat testing using a modified \texttt{rc11+lb.cat} model to show that all of the above behaviours disappear when load-to-store ordering is permitted.

Many differences in Tab.~\ref{fig:results} arise from data races. The C/C++ model flags data races as undefined behaviour, and we ignore false positives on that basis. Of course, we assume the models are correct, which is a limitation we accept given these promising results.

\subsection{Limitations of Model-based Testing}\label{limit}
Our technique has three limitations: model correctness, model completeness, and simulation scalability. When exploring each case we found new bugs detailed below.

We assume the source and target models are correct. We found a bug in the (unofficial) Armv7 model~\cite{aarch32} that the state-of-the-art techniques miss. We found a bug when compiling for the Armv7-a architecture using a \textit{Store Buffering} litmus test. The outcome of the test was allowed by the unofficial Armv7 model~\cite{aarch32}, but is forbidden by the RC11 model~\cite{simonrc11} and the Armv7 hardware we checked. The problem was that the Armv7 model was allowing accesses to be reordered when it should have been forbidden. We reported the bug and fixed the model~\cite{dmbish}. As the state-of-the-art (\textsection\ref{sota}) depends only on source models this limitation is unique to \telechat{}.

Next, we assume that models support language features under test. We reported~\cite{constbug} a bug in the implementation of 128-bit \texttt{const} atomic loads. We found that \texttt{const} was miscompiled when loading constant atomic data - it crashes at run-time as the C/C++ load is implemented using a store instruction that attempts to write to read-only memory. Simulation under the Arm AArch64 model~\cite{aarch64} will miss this bug, as \texttt{const} read-only memory is unsupported, and so we augment the model to flag \texttt{const} violations. Whilst conducting this study a fix was proposed in LLVM by engineers~\cite{applefix}, but the problem remains as the fix only applies to Armv8.4 or above (similar code exists for Intel x86-64, RISC-V, IBM PowerPC backends). Upon discussing this bug with Arm's compiler engineers, we conclude there is no (lock-free) fix for Armv8.0, since the 128-bit load instruction is not guaranteed to be single-copy atomic unless the Armv8.1 Large Systems Extension is implemented.

Lastly, the state explosion problem limits the bounds of what \texttt{herd} can test. Consider Fig.~\ref{lb_3rc11} that extends Fig.~\ref{lb_c4} with an additional thread \texttt{P2}. If Fig.~\ref{lb_3rc11} is compiled and simulated under the Arm AArch64 model~\cite{aarch64}, then \texttt{herd} does not terminate with a one hour timeout. Since \texttt{herd} enumerates executions, it suffers from the state-explosion problem. Without optimisation, execution time of assembly litmus tests expands factorially as the test size increases, practically limiting \texttt{herd}'s ability to scale much above programs of the order of 40-50 lines of code.

\begin{figure}[h]
  \centering
  \begin{tabular}{c}
\begin{lstlisting}[language=C, style=mystyle]
{ *x = 0, *y = 0 }

void P0 (int* y,int* x) {
  int r0 = *x;
  atomic_thread_fence(memory_order_relaxed);
  *y = 1;
}
void P1 (int* z,int* y) {
  int r0 = *y;
  atomic_thread_fence(memory_order_relaxed);
  *z = 1;
}
void P2 (int* z,int* x) {
  int r0 = *z;
  atomic_thread_fence(memory_order_relaxed);
  *x = 1;
}
exists (P0:r0=1 /\ P1:r0=1 /\ P2:r0=1)
\end{lstlisting}
  \end{tabular}
  \caption{A C/C++ litmus test, when compiled targeting Arm AArch64 does not terminate under simulation.}
  \label{lb_3rc11}
\end{figure}

We sidestep the state explosion problem by optimising compiled litmus tests. The problem depends on the size of compiled program executions. Whilst \texttt{herd} considers \texttt{int r0 = *x} to be one load of $x$ and one store to \texttt{r0}, the compiled program uses many instructions. For every C/C++ access in the source program, LLVM or GCC generates at least three Arm assembly instructions: \texttt{ADRP} to calculate the pointer to $x$, a \texttt{LDR} to load the location $x$ into a register, and a \texttt{LDR} to load the value of $x$. As each instruction generates multiple loads or stores, the number of events in target executions is an order of magnitude larger than executions of the source. Computing whether such a graph is allowed using \texttt{herd} induces a state explosion as each \texttt{LDR} contributes to the reads-from relation (def.~\ref{exe}). We optimise \texttt{ADRP *x;LDR;LDR/STR x $\rightsquigarrow$ LDR/STR x} sequences in \telechat{}, and contribute a suite of similar optimisations for each architecture we test. Using \telechat{}, simulating the compiled Fig.~\ref{lb_3rc11} terminates in milliseconds. Checking the soundness of our optimisations is future work, but an informal argument is that the \texttt{herd} simulator uses symbolic locations. The locations associated with accesses we remove cannot be named by other threads and an access cannot side-effect other symbolic locations. Soundness depends on the non-interference of other threads after applying our optimisations. Scalability is a still problem in theory, but in practice we only see timeouts with large (5+ threads or 6+ shared variables) tests.

We end by discussing our working hypothesis. Many authors~\cite{cmmtest,10.1145/3563292,inbook} claim simulation is unlikely to scale. For instance, Morisset et al. claim~\cite{cmmtest}: \textit{``[herd] is unlikely to scale to the complexity of hunting C11/C++11 compiler bugs.''} There is however a decade of evidence to suggest small (two threads at around twenty LoC) litmus tests are effective in finding bugs in hardware implementations~\cite{Alglave:2011:LRT:1987389.1987395}, Arm hardware designs~\cite{10.1145/3466752.3480087}, GPUs~\cite{10.1145/2694344.2694391}, the Linux Kernel~\cite{Alglave:2018:FSC:3173162.3177156}, C/C++~\cite{Vafeiadis:2015:CCO:2676726.2676995,Lahav:2017:RSC:3062341.3062352}, and more. Since every concurrency compilation bug we know of can be demonstrated by a small litmus test, we question whether simulation \textit{needs} to scale. We expect the \textit{small-scope hypothesis}~\cite{10.5555/3031843.3031849} holds: ``that a high proportion of errors can be found by testing a program for all test inputs within some small scope''. Whether there are bugs that are triggered by large-programs \textit{only} is an area for future work. 

\subsection{Industry Impact}\label{imp}
We improved compilers - used in industry - in two ways. We answered queries from Arm's partners~\cite{LDAPR} and deployed automated regression testing for Arm Compiler. \telechat{} is the first tool of its kind to be deployed in industry. 

We assisted Arm's engineers with a query from Google~\cite{LDAPR} engineers. Following compelling performance metrics on hardware, Google's engineers proposed to change the implementation of C/C++ acquire loads to use the \texttt{LDAPR} instruction instead of \texttt{LDAR} when the Armv8.3-a weak release consistency extension is enabled. The \texttt{LDAPR} instruction allows more re-orderings than \texttt{LDAR}, however experts failed to find a bug under this proposal. Reviewers were inclined to accept the proposal without a correctness proof, but the proof was estimated to take three months. With \telechat{}, we provided experimental testing of the proposal and Arm's compiler team chose to accept the proposal based on our work~\cite{LDAPR}.

Lastly, we deployed automatic regression testing of Arm Compiler. Arm's compiler teams wish to test whether Arm Compiler correctly translates concurrent C/C++ programs targeting the Armv7 and Armv8 architectures. We conduct differential testing of Arm Compiler and deployed \telechat{} in their automated testing infrastructure using an artefact like the one we provide with this work. As far as we know, \telechat{} is the first compiler testing tool (for concurrency) to be deployed in a production setting - it is an industry first.

\section{Conclusions and Further Work}\label{conc}
We present the \telechat{} automated compiler testing technique for programs with relaxed memory concurrency semantics. We documented its design (\textsection\ref{technique}) and implementation (\textsection\ref{im}) with a reproducible artefact. We show how it improves on the state-of-the-art (\textsection\ref{c4}) and the real-world benefits~\cite{LDAPR} \telechat{} brings when assisting Arm's compiler team. We refute a claim~\cite{cmmtest} made by the state-of-the-art whilst exploring a novel kind of concurrency bug (\textsection\ref{local}) that evaded detection until now. We conducted large-scale differential testing of LLVM and GCC (\textsection\ref{large}), which is the most extensive concurrency test campaign to date as far as we know. We assisted Arm's compiler team with two queries~\cite{LDAPR} from Arm's partners, reported four new bugs~\cite{ldpbug,constbug,stp,swp}, and found one behaviour known by concurrency experts but missed by the state-of-the-art (Fig.~\ref{lb_c4}). We fixed a bug in the Armv7 model~\cite{dmbish}, and generated several new lines of inquiry whilst exploring the limitations of our work (\textsection\ref{limit}). Lastly, \telechat{} is, as far as we know, the industry's first concurrency tool to be deployed in automated testing for Arm Compiler.

We sought a practical bug (def.~\ref{fault}) finding technique for production compilers. By using official architecture models, we achieve this goal, but defer a soundness proof to future work. Proving soundness requires a model of ground truth~\cite{10.1109/TSE.2014.2372785}. Such a model changes with business pressures; thus fixing a ground truth is a challenge. Even the official Armv8 model~\cite{aarch64} forbids many behaviours of the older Arm model as no implementer has built a machine that exploited its additional relaxations\footnote{The Armv8 model is experimentally stronger~\cite{exp} than the Armv7 model.}. The C/C++ model sees similar evolutions~\cite{Vafeiadis:2015:CCO:2676726.2676995,10.1145/2837614.2837637,Lahav:2017:RSC:3062341.3062352,10.1145/1375581.1375591,Batty:PhD} and challenges for compiler engineers to avoid out-of-thin-air behaviours. For soundness to hold as compilers and models are updated, automating model-based proof~\cite{10.1145/3290382,Chakraborty:2019:GTR:3302515.3290383,10.1007/978-3-319-89884-1_36} for production compilers is desirable. In the absence of repeatable proof, \telechat{} provides practical testing.

Test generation is an area for future work. Fig.~\ref{fig:counter} induces two bugs in past versions of LLVM and GCC. We expect that exploring the state-space of litmus tests or conducting mutation-based testing~\cite{10.1109/ASE.2013.6693142} will find more bugs. The Alive2~\cite{10.1145/3453483.3454030} tool finds bugs whilst exploring the state space of sequential tests; but it is unknown whether it scales to concurrent programs in light of the state explosion (\textsection\ref{limit}) problem. Windsor et al.~\cite{windsor2022high} conduct metamorphic testing of LLVM, but miss the bugs we report~\cite{stp,constbug,ldpbug} - we expect there are more bugs out there. We just need the tests to find them.

We present new lines of inquiry for compiler testing. The field of testing with (C/C++) models is over a decade old - we feared little progress could be made. Indeed, recent work~\cite{DBLP:conf/asplos/BeckBSCBFV0H23,10.1145/3519939.3523719,10.1145/2994593} focuses instead on the \textit{porting problem}. We uncovered new lines of inquiry that suggest there is still work to be done. One such line - inspired our \texttt{const} work (\textsection\ref{limit}) - is studying the interaction between sequential and concurrent C/C++. \texttt{const} \texttt{atomic} loads induce run-time crashes~\cite{constbug}, but it is unclear \textit{how} such types are used in practice. The call for clarity on the compilation of atomics increases as multi-core machines play an increasing role in our lives.

\section*{Acknowledgment}
We thank supervisors James Brotherston and Earl Barr. Luc Maranget, Ana Farinha, Alastair Donaldson, John Wickerson, Tyler Sorensen, Shale Xiong, Alastair Reid, Peter Smith, Wilco Dijkstra, Arm’s Compiler Teams and Arm Architecture \& Technology Group for their feedback and assistance. This work was supported by the Engineering and Physical Sciences Research Council [grant number EP/V519625/1]. The views of the authors expressed in this paper are not endorsed by Arm or any other company mentioned.

\section*{Artefact Appendix}

\subsection{Abstract}

The artefact consists of the \telechat{} tool and scripts provided with this paper. \telechat{} builds on the herd tool-suite~\cite{herdtools} and its models. As such the results are liable to change. We acquired all badges. For comments please contact \url{luke.geeson@cs.ucl.ac.uk}.

\subsection{Artefact Checklist}

\begin{enumerate}
\item \textbf{Algorithm:} \telechat{}.
\item \textbf{Program:} \texttt{l2c,c2s,s2l} and herdtools~\cite{herdtools}.
\item \textbf{Compilation:} Includes LLVM 11, GCC 9.2, GCC 10.
\item \textbf{Models:} From herd toolsuite~\cite{herdtools}.
\item \textbf{Data Set:} Tests generated using provided \texttt{c11.conf}.
\item \textbf{Test Environment/Binary:} Docker Ubuntu 20.04.
\item \textbf{Hardware:} Either x86-64 or Arm AArch64 machines.
\item \textbf{Run-time State:} not sensitive to run-time state.
\item \textbf{Metrics:} Outcomes of executing tests under models.
\item \textbf{Output:} Console and \texttt{.log} files.
\item \textbf{Experiments:} Makefile provided reproduces results.
\item \textbf{Disk-space requirements:} 5GB for Docker image, +100GB for the large-scale study (\textsection\ref{large}).
\item \textbf{Time needed to prepare workflow}: Everything is ready.
\item \textbf{Time needed to complete experiments}: $\sim10$ hours.
\item \textbf{Licenses:} CeCILL-B license.
\item \textbf{Workflow Frameworks:} Makefile, GNU Parallel~\cite{Tange2011a}.
\item\textbf{Archived(DOI):} \url{https://doi.org/10.5281/zenodo.10204529}
\item \textbf{Available:} Zenodo or Docker Hub\footnote{\url{https://hub.docker.com/r/lukeg101/telechat-artefact/tags}}.
\end{enumerate}

\subsection{Description}

\subsubsection{How Delivered}
The artefact is available on Zenodo and consists of a Docker container with the \telechat{} tool, compilers under test, and scripts required to reproduce results.

\subsubsection{Hardware Dependencies}
Either an Intel x86-64 or Arm AArch64 based machine. The artefact was tested using a MacBook Pro with a dual-core Intel i7 CPU, a Lenovo P720 with 2xIntel Xeon Gold 5120T CPUs (56 cores), a MacBook Air with an 8-core Apple M1 (Arm AArch64), a Cavium Thunder X2 with 2x28-core CPUs (Arm AArch64), and under x86-64 emulation (using the M1 machine). 

\subsubsection{Software Dependencies}
\telechat{} requires a Linux distribution such as Ubuntu. Including:
\begin{itemize}
\item The C/C++ compiler under test (multi-lib cross-compilers work best on multiple platforms).
\item GNU binutils, e.g. \texttt{binutils-riscv64-linux-gnu}.
\item GNU Parallel~\cite{Tange2011a}, libxml2, time, and libc6
\end{itemize}

\subsection{Installation} \label{install}
\begin{enumerate}
\item Download and install Docker. For example on Ubuntu 20.04 you can install docker using the official guide\footnote{\url{https://docs.docker.com/desktop/install/ubuntu/}}
\item Download \texttt{telechat-artefact-arch.tar.gz} from Zenodo (where \texttt{arch} is either \texttt{arm64} or \texttt{x86}).
\item Load the Docker container:
\begin{verbatim}
> docker load -i \
  telechat-artefact-arch.tar.gz
\end{verbatim}
\item Run the Image:
\begin{verbatim}
> docker run -it \
  lukeg101/telechat-artefact
\end{verbatim}
This runs the Ubuntu image and mounts the current directory into the container at \texttt{artefact-output}.
\end{enumerate}

Alternatively, if you wish to install from Docker Hub, we provide Intel x86-64 and Arm AArch64 builds:
\begin{verbatim}
> docker pull \
  lukeg101/telechat-artefact:latest
\end{verbatim}
Then run:
\begin{verbatim}
> docker run -it \
  lukeg101/telechat-artefact:latest
\end{verbatim}

\subsection{Experiment Workflow}
A \texttt{Makefile} drives the \telechat{} toolchain, and examples of how to use it are provided in the \texttt{README.md}. For example, to run the ``smoketest'' in the docker container, type:
\begin{verbatim}
artefact> make examples
\end{verbatim}
The Readme contains instructions on how to customise testing and generate different test benchmarks.

\subsection{Paper Claims}
\begin{enumerate}
\item Fig. 7 has outcomes in Fig. 8 (left). under the RC11 model~\cite{Lahav:2017:RSC:3062341.3062352}, when compiled for Arm AArch64, it has the Fig.8 (right) outcomes.
\item Windsor et. al miss~\cite{windsor2022high} miss the load buffering behaviour of Fig.7. \telechat{} observes it.
\item We exercise all the features in Table III. when testing LLVM and GCC for the architectures listed.
\item We get the results in Table IV under the RC11 model~\cite{Lahav:2017:RSC:3062341.3062352}, but if we permit load-to-store reordering all positive differences disappear.
\item Compiling and Optimising Fig.11 using \telechat{} enables its simulation to terminate in milliseconds.
\end{enumerate}
A number of minor claims appear in the paper, like how we added a vector datatype to \texttt{herd}. To keep this appendix small we refer the reader to \telechat{} generated tests that use these features. To validate the bug reports, please see our bug board\footnote{\url{https://lukegeeson.com/blog/2023-10-17-Telechat-Bug-Board/}}

\subsection{Evaluation and Expected Results}
We assume you are running with a clean directory.\\

\paragraph*{\textbf{Claim 1 ($<$ 5 minutes on an Apple M1 machine)}}\hfill

Please run:
\begin{verbatim}
artefact> make examples
\end{verbatim}
Check the log:
\begin{verbatim}
artefact> cat artefact-output/Output/logs\
  /examples_int_C_tests_llvm-O3-AArch64_\
  mcompare.log
\end{verbatim}
The source and compiled program outcomes are tabulated, \texttt{LB004\_examples\_int\_C\_tests} has new behaviour:
\begin{verbatim}
c11_[...]_tests   a64_[...]_tests
[0:r0=0; 1:r0=0;] +[P0_r0=1; P1_r0=1;]
[0:r0=0; 1:r0=1;]
[0:r0=1; 1:r0=0;]
\end{verbatim}\\

\paragraph*{\textbf{Claim 2 ($<$ 1 minute checking manually)}}\hfill

Windsor et. al~\cite{windsor2022high} state: \begin{quote}
``\textit{we experimented with using the stronger RC11 memory model of Lahav et al.~\cite{Lahav:2017:RSC:3062341.3062352} as the input to our test-case generator, RMEM [a simulator parameterised over architecture models, not part of C4] identified as a bug the ‘load buffering’ test that RC11 forbids, but C11 and AArch64 permit.}''
\end{quote}

We observe load buffering (Figs.7+8) in \textbf{Claim 1}.\\

\paragraph*{\textbf{Claim 3 ($\sim$ 10 hours on a 224 core ThunderX2)}}\hfill

Please run:
\begin{verbatim}
artefact> make all CONF_FILE=c11.conf
\end{verbatim}
\textit{Warning}: This requires a powerful machine to run.\\

Once done, the \texttt{Output} directory should reveal tests that contain the following:
\texttt{fence}, \texttt{*x}, \texttt{if}, \texttt{atomic\_load}, \texttt{atomic\_store}, \texttt{clang-11}, \texttt{gcc-10}, \texttt{-O1}, \texttt{-O2},\texttt{-O3}, \texttt{-march=armv7}, \texttt{-march=x86-64}, \texttt{-march=mips64}, \texttt{powerpc-linux-gnu}, \texttt{aarch64-linux-gnu}, \texttt{riscv64-pclinux-gnu}, and so on\ldots\\

\paragraph*{\textbf{Claim 4 ($\sim$ 10 hours on a 224 core ThunderX2)}}\hfill

Please run:
\begin{verbatim}
artefact> make all CONF_FILE=c11.conf
\end{verbatim}

Once done, the numbers in Table IV should match the \textit{+ve} and \textit{-ve} differences listed on the console output. Since the C/C++ standards permit load-to-store re-ordering (ie load buffering), observe that all of the \textit{+ve} differences go away when we use the \texttt{rc11+lb.cat} model:

\begin{verbatim}
artefact> make all CONF_FILE=c11.conf \
  CMEM=rc11+lb.cat
\end{verbatim}
\textit{Warning}: This requires a powerful machine to run.\\

\paragraph*{\textbf{Claim 5 ($<$ 5 minutes on an Apple M1 machine)}}\hfill
Please run:
\begin{verbatim}
artefact> make examples
\end{verbatim}
And then you can see the compiled (and optimised) Fig.10:

\begin{verbatim}
artefact> cat artefact-output/Output\
  /examples_int_C_tests/tgt/llvm-O3-AArch64\
  /3.LB004_examples_int_C_tests.litmus
\end{verbatim}

Simulation timings are in the \texttt{herd} log:
\begin{verbatim}
artefact> cat artefact-output/Output \
  /examples_int_C_tests/tgt/      \
  llvm-O3-AArch64/all_a64_llvm-O3-\
  AArch64_examples_int_C_tests.log
\end{verbatim}

Observe that simulation took $\sim$3 milliseconds (subject to your CPU clock speed and memory latency):
\begin{verbatim}
Test 3.LB004_examples_int_C_tests Allowed
States 8
[P0_r0]=0; [P1_r0]=0; [P2_r0]=0;
[P0_r0]=0; [P1_r0]=0; [P2_r0]=1;
[...]
Time 3.LB004_examples_int_C_tests 0.03
\end{verbatim}

On the other hand, Consider the \texttt{unoptimised.litmus} test, adapted from LLVM-11 code taken from godbolt.org\footnote{\url{https://godbolt.org/z/G9b4Pq1YK}} (which is the same as \texttt{3.LB004}) that we have not seen terminate after running for 1 hour on an Apple M1 machine:
\begin{verbatim}
artefact> make dnf
\end{verbatim}
\textit{Warning: It is unclear whether herd terminates with this input}

This motivates our need to optimise compiled litmus tests.

\subsection{Experiment Customisation}

You can customise the experiments when invoking Make:
\begin{itemize}
\item Generate different C/C++ tests using a config file (default: None, options: \texttt{c11.conf}, \texttt{c11\_acq.conf}):
\begin{verbatim}
artefact> make examples \
  CONF_FILE=c11.conf
\end{verbatim}
\item Set source model (default \texttt{rc11.cat}, options: \texttt{c11\_partialSC.cat},\texttt{c11\_simp.cat},\texttt{rc11.cat}, \texttt{rc11+lb.cat}):
\begin{verbatim}
artefact> make examples \
  CMEM=c11_simp.cat
\end{verbatim}
\item Set simulation timeout, (default \texttt{120.0} seconds):
\begin{verbatim}
artefact> make examples TIMEOUT=1.0
\end{verbatim}
\item Test other compilers. Outside the container, add a profile to \texttt{profiles.json}, add the profile name (such as \texttt{llvm-O3-AArch64}) to the \texttt{PROFILE} variable in the \texttt{Makefile}, add the \texttt{MODEL\_}\textit{profile} to the \texttt{Makefile}, and re-run \texttt{./build.sh \&\& ./run.sh}.
\end{itemize}

\subsection{Available Benchmarks}
The benchmarks used can be generated by providing a \texttt{CONF\_FILE} parameter to the \texttt{Makefile}:
\begin{itemize}
\item \textsection IV.D: \texttt{c11.conf}: for the large-scale differential testing 
\item \textsection IV.F: \texttt{c11\_acq.conf}: for the LDAPR case study 
\end{itemize}

This article represents a personal opinion that is not endorsed by Arm.

\balance
\bibliographystyle{acm}
\bibliography{paper}

\end{document}